\documentclass[aps,twocolumn,showpacs]{revtex4}

\usepackage{graphicx}
\usepackage{lipsum}
\begin{document}

\title{Thermodynamics of a time dependent and dissipative oval billiard: a 
heat transfer and billiard approach}

\author{Edson D.\ Leonel}

\affiliation{Departamento de F\'isica, UNESP - Univ. Estadual Paulista - Av. 24A, 1515 - 
Bela Vista - 13506-900 - Rio Claro - SP - Brazil \\
Abdus Salam International Center for Theoretical Physics, Strada  Costiera 
11, 34151 Trieste, Italy}

\author{Marcus Vin\'icius Camillo Galia}

\affiliation{Departamento de F\'isica, UNESP - Univ. Estadual Paulista - Av. 24A, 1515 - 
Bela Vista - 13506-900 - Rio Claro - SP - Brazil}

\author{ Luiz Antonio Barreiro}

\affiliation{Departamento de F\'isica, UNESP - Univ. Estadual Paulista - Av. 24A, 1515 - 
Bela Vista - 13506-900 - Rio Claro - SP - Brazil}

\author{Diego F. M. Oliveira}

\affiliation{Center for Complex Networks and Systems Research, School of Informatics and 
Computing, Indiana University, Bloomington - Indiana, USA. \\
Department of Chemical and Biological Engineering, Northwestern University, Evanston, Illinois 60208, USA.\\
Northwestern Institute on Complex Systems (NICO), Northwestern University, Evanston, Illinois 60208, USA.}

\date{\today} \widetext


\begin{abstract}
We study some statistical properties for the behavior of the average squared 
velocity -- hence the temperature -- for an ensemble of classical particles 
moving in a billiard whose boundary is time dependent. We assume the collisions 
of the particles with the boundary of the billiard are inelastic leading the 
average squared velocity to reach a steady state dynamics for large enough 
time. The description of the stationary state is made by using two different 
approaches: (i) heat transfer motivated by the Fourier law and; (ii) billiard 
dynamics using either numerical simulations and theoretical description.
\end{abstract}

\maketitle

\section{Introduction}
\label{sec1}

The initial mark in the investigation of billiards theory is related to Birkhoff 
\cite{ref1} in beginning of last century. Since then this research area has 
developed significantly. Birkoff considered the investigation of the motion of a 
free point-like particle -- representing a billiard ball -- in a bounded 
manifold. However the modern investigations of billiards are indeed related to 
the results of Sinai \cite{ref3,ref3a} and Bunimovich \cite{ref4,add1} who made 
rigorous demonstrations on the topic.

A billiard is a dynamical system composed of a particle, or an ensemble of 
non interacting particles, moving confined to a domain with a piecewise-smooth 
boundary \cite{ref5} where they collide. The specular reflections occur under 
the condition the boundary is sufficiently smooth. In such case the tangent 
component of the velocity of the particle measured with respect to the border 
where collision happened is unchanged while the normal component reverses sign. 
There are many results nowadays considering either static \cite{ref6,ref7,ref8,ref8a,ref8b,ref8c,ref8d,ref8e,ref8f}
and time dependent boundaries \cite{td2,td3,ref5a,ref5b}. A phenomenon well known in 
time dependent boundary is the Fermi acceleration \cite{ref9} as well 
as the so called Loskutov-Ryabov-Akinshin (LRA) conjecture \cite{ref10,ref11}. 
Fermi acceleration \cite{ref9} is a phenomenon where an ensemble of particles 
acquires unlimited energy from collisions with an infinitely heavy and moving  
wall. The conjecture itself claims that if chaos is present in the dynamics of 
a particle in a static version of the billiard, then this is a sufficient -- 
but not necessary -- condition to observe Fermi acceleration when a time 
perturbation to the boundary is introduced. Many different billiards exhibit 
Fermi acceleration under time perturbation to the boundary including the Lorentz 
gas \cite{ref12,ref12a}, oval billiard \cite{ref13}, stadium \cite{ref14} and other 
shapes \cite{ref17}. The elliptical billiard is an exception and the LRA 
conjecture does not apply to it. For the static boundary, the elliptical 
billiard is integrable \cite{ref5} and the phase space is composed of rotating 
and librating orbits. A curve which separates these two different regimes is 
called as separatrix. Lenz et al. \cite{ref19,ref20} shown that when the 
boundary of the billiard is allowed to be time dependent, the separatix curve 
is  replaced by a stochastic layer allowing cross visitations 
from regions of libration and rotation. For the static case both 
energy, $E$, and angular momenta about the two foci, $F$, are constants of 
motion \cite{ref21}, leading the system to be integrable. However, when time 
perturbation is introduced, the observable $F$ (see Ref. \cite{ref19}) 
experiences strong and fast variations from the crossing of orbits coming 
from rotation and those leaving from libration. The successive crossings 
produce the stochasticity required in the LRA conjecture, hence leading the 
time dependent elliptical billiard to exhibit Fermi acceleration. This result 
is considered a counter example of the LRA conjecture. Latter on, 
investigations on different models have proved the Fermi acceleration is not a 
robust phenomenon since a very small amount of dissipation is enough to suppress 
the phenomenon \cite{ref22}. Consideration of inelastic collisions in the 
elliptical billiard \cite{ref23} has proved the successive crossings of orbits 
coming from rotation region and entering libration domain -- and vice versa -- 
are interrupted suppressing the Fermi acceleration.

The motion of the time dependent boundary can be related to a more physical 
situation. Due to the thermal fluctuations the position of each atom 
that compose the boundary is allowed to move locally. Such oscillation of 
the atoms, and hence of the boundary, can be brought to the context of billiard 
which allows connections of the observables obtained from the velocity of the 
particle -- hence the kinetic energy -- to the thermodynamics, particularly 
the temperature and entropy.

In this paper we investigate some dynamical properties for an ensemble of 
particles confined in an oval billiard whose boundary is moving in time. Our 
main goal is to understand and describe the dynamics of the average squared 
velocity for a gas of noninteracting particles. We will do this by using two 
different procedures. Because the boundary of the billiard is moving, as soon 
as the particles collide, there is a change of energy of the particle. 
Therefore, the first procedure considered involves the heat transfer Fourier 
equation. We write and solve the Fourier equation considering the geometrical 
properties of the boundary. The resulting equation is that the temperature of 
the gas settles down for sufficiently long time as the temperature of the 
boundary, hence the average squared velocity, reaches the thermal equilibrium. 
The second one involves the formalism commonly used in billiard problems. We 
write down the equations of the mapping that describe the dynamics of the 
problem and extract some average properties for the squared velocity of the 
particles. The properties are obtained either by straightforward numerical 
simulations as well as analytically. The results obtained on the analytical 
approach are remarkable well fitted by the numerical simulations. The first 
approach however uses the time as the dynamical variable while the second one 
uses the number of collisions of the particles with the boundary. The two 
dynamical variables are not trivially connected among themselves. Therefore, by 
the use of an empirical function, we find a straight relation between these two 
parameters.

This paper is organized as follows. In Sec. \ref{sec2} we discuss the 
properties of the average squared velocity by the use of heat transfer Fourier 
equation. We use some geometrical properties of the boundary to fit into the 
required parameters of the equation. Section \ref{sec3} is devoted to construct 
the billiards approach of the problem. We them obtain the equations that 
describe the dynamics of the model and discuss the several types of 
characterization including steady state, dynamical regime, numerical 
simulations, critical exponents and the behavior of the probability 
distribution function for the velocity of the particles. The connection of the 
two parts is made in Sec. \ref{sec4} where a relation between the time and 
number of collisions is obtained. Conclusions and discussions are made in Sec. 
\ref{sec5}.

\section{Heat transfer approach}
\label{sec2}

\begin{figure}[htb]
\centerline{\includegraphics[width=1\linewidth]{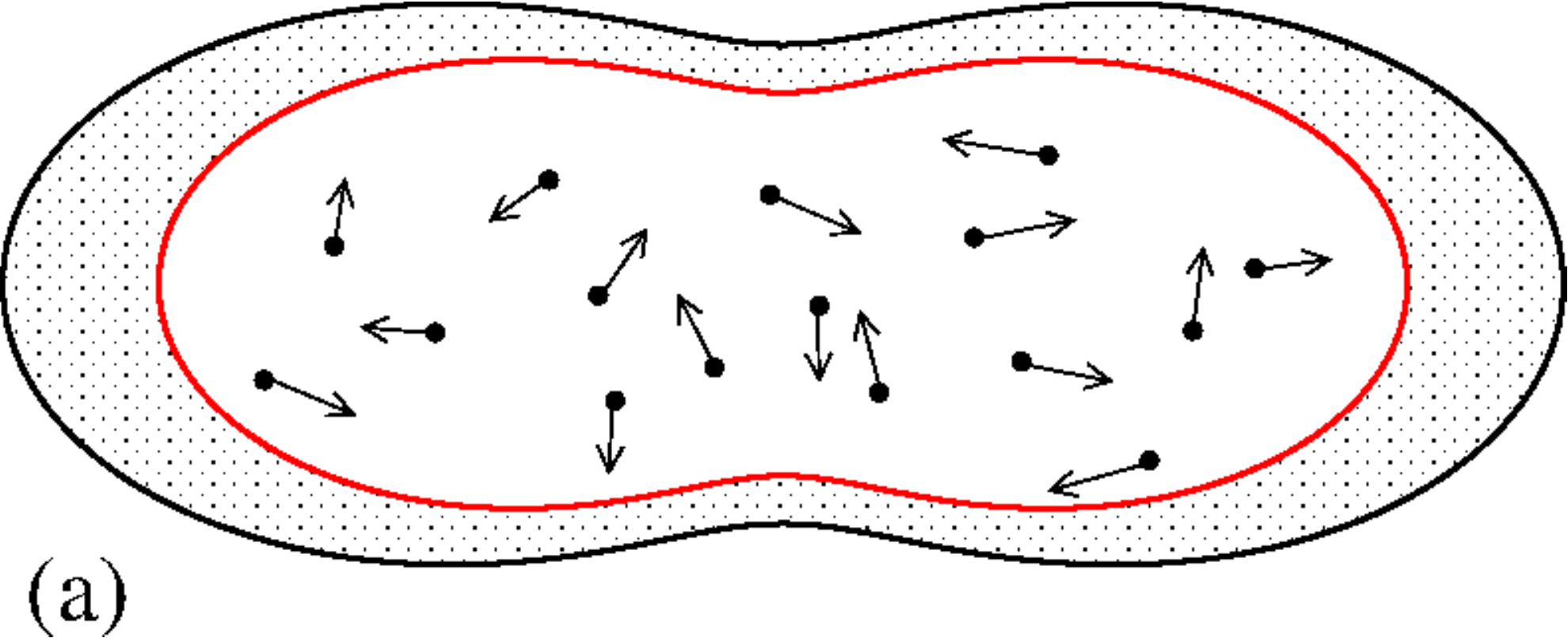}}
\centerline{\includegraphics[width=1\linewidth]{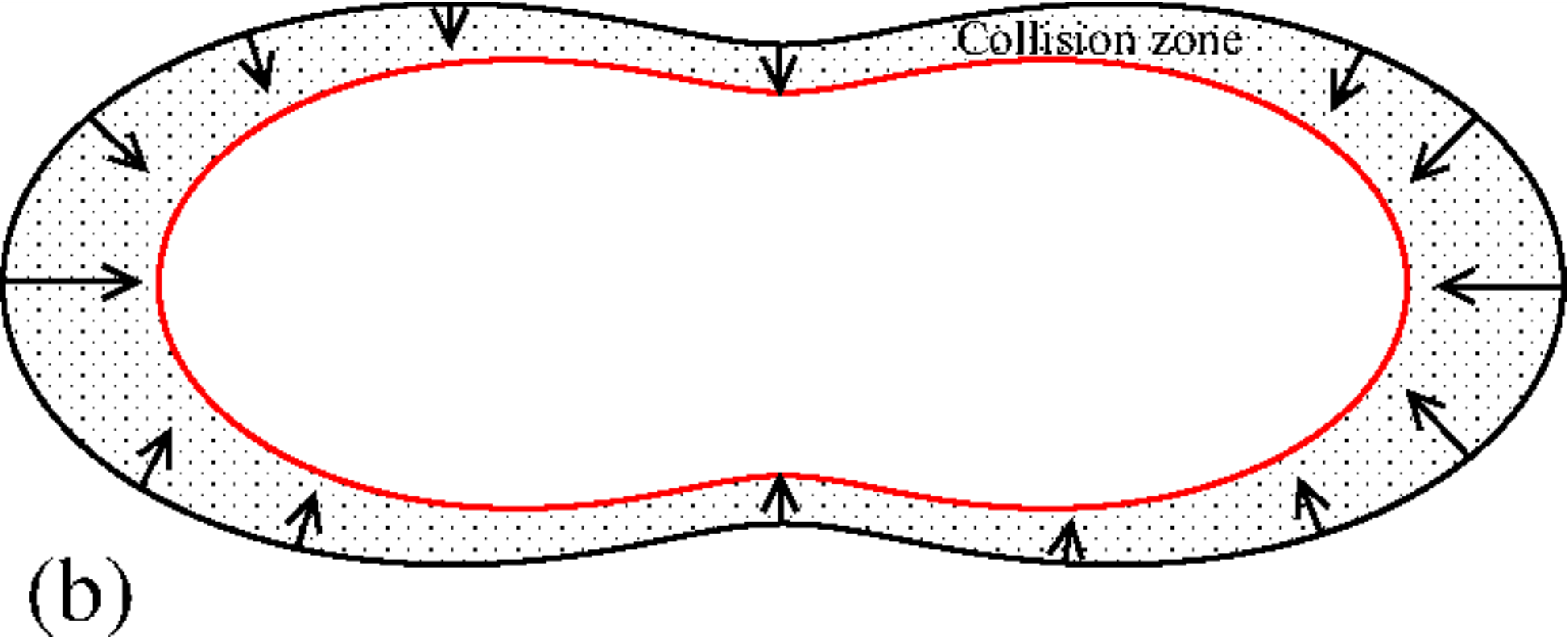}}
\caption{(a) Sketch of the billiard boundary and an ensemble of noninteracting 
particles. (b) Illustration of the heat transfer region. The arrows direction 
point the heat flux.}
\label{Fig1}
\end{figure}

We discuss in this section the approach involving heat transfer. To start with 
we assume that there is a set of identical particles moving inside a closed boundary. 
The density of the particles is considered sufficiently small such that the 
particles are noninteracting. Figure \ref{Fig1}(a) shows an illustration of 
the system. We assume the boundary of the billiard is moving in time, therefore, 
this is the mechanism responsible for the exchange of energy with the 
particles: collisions! 
The boundary is at a temperature $T_b$ that is considered fixed and does not 
change with the dynamics of the particles. Hence the boundary works as a 
thermal bath and two obvious conclusions can be extracted. If the temperature 
of the gas of particles is less than $T_b$, then the boundary gives energy to 
the particles raising up the temperature of the gas. On the other hand, if the 
temperature of the particles is larger than $T_b$, the heat bath absorbs energy 
from the particles and dissipate it along with the chain of nearby atoms of the 
boundary -- hence reducing the temperature of the gas. There is a region near 
the border of the billiard where the particles can exchange energy which we 
denote as a collision region.

The Hamiltonian that describes the dynamics of each particle is given by
\begin{equation}
H={{p^2}\over{2m}}+V(q_x,q_y,t),
\label{eq1}
\end{equation}
where $p^2=p_x^2+p_y^2$ corresponds to the momentum of the particle and $V$ is 
the potential energy which is written as
\begin{equation}
V(q_x,q_y,t)=\left\{\begin{array}{ll}
0~~{\rm if}~~(q_x,q_y,t)<R(t)\\
\infty~~{\rm if}~~(q_x,q_y,t)=R(t)\\
\end{array}
\right.,
\label{eq2}
\end{equation}
where $R$ is the radius of the boundary written in polar coordinates, which 
assumes the following form in this work $R(\theta,\eta,t)=1+\eta 
f(t)\cos(p\theta)$, where $p$ is any integer number. A non integer number leads 
to an open boundary to where the particles can escape. $\eta$ is a parameter 
which controls the circle perturbation. If $\eta=0$, the boundary is a circle, 
that is integrable \cite{ref5}, in billiards terminology, due to the 
conservation of energy and angular momentum, while $\eta\neq 0$ leads the phase 
space to be mixed when $f(t)$ is a constant \cite{ref21}. The function $f(t)$ 
leads to the time perturbation of the boundary and we consider in this work two 
different types of perturbation: (i) periodic oscillations and; (ii) random 
oscillations. For case (i) the function is written as 
$f(t)=1+\varepsilon\cos(\omega t)$ where $\varepsilon$ is the amplitude of 
oscillation and $\omega$ is the angular frequency, which we set it as fixed 
$\omega=1$. For the random case (ii), the function $f(t)$ assumes the same 
expression as in the case (i) however, at the instant of the impact, we assume 
there is a random phase $Z(t)$, given random numbers $Z\in[0,2\pi]$, such that 
the velocity of the moving boundary is given by 
$\vec{V}_b(t)={{d}\over{dt}}[\vec{R}_b(t)+Z(t)]$. This choice is made in such a 
way to avoid the possibility of having the chance of the particles moving 
outside of the boundary, hence a non physical situation. At the same time, this 
is an easy way to introduce randomness in the model. In this section we discuss 
the 
thermodynamical properties based on the heat transfer equation -- Fourier law -- 
and the geometrical parameters of the boundary will be used in the approach. In 
next section we describe the dynamics by using the billiards formalism hence 
writing the dynamical equations of the particle and averaging the velocity as a 
function of the number of collisions as well as along an ensemble of particles.

The equation that governs the heat transference \cite{blu} is written as
\begin{equation}
{{\partial Q}\over{\partial t}}=-\kappa \ell{{\partial T}\over{\partial x}},
\label{eq3}
\end{equation}
where $\kappa$ corresponds to the heat conductivity coefficient, $\ell$ is 
length along the boundary to where the heat can flow and is obtained from the 
geometrical properties of the boundary, ${{\partial Q}\over{\partial t}}$ 
denotes the flux of heat from a region where there is a temperature difference 
$\Delta T$ and ${{\partial T}\over{\partial x}}$ corresponds to the temperature 
gradient. We present a short discussion of the Fourier equation in Appendix 1 
and an interpretation of the heat conductivity coefficient $\kappa$ for the 
one-dimensional case. The minus $(-)$ is related to the fact the heat flow from 
the region of higher to the lower temperature, hence opposite to the temperature 
gradient \cite{blu}. Figure \ref{Fig1}(b) illustrates schematically the 
collision zone and the region to where heat can flow. The effective length 
$\ell$ to where heat can flow is obtained from 
$\ell=\int_0^{2\pi}R(\theta,\eta,\varepsilon,p,t)d\theta=\int_0^{2\pi}[1+\eta[
1+\varepsilon\cos(t)]\cos(p\theta)]d\theta=2\pi$.

The two steps we consider to solve Eq. (\ref{eq3}) is to obtain the 
corresponding  expressions for: (i) ${{\partial Q}\over{\partial t}}$ and; (ii)
${{\partial T}\over{\partial x}}$ in such a way that its solution can be 
obtained. We know 
that the density of particles is considered sufficiently small so that each 
particle does not interact with any other. Therefore, the energy of each 
particle is due to the energy associated to the state of its motion, hence 
kinetic energy. From the energy equipartition theorem we have that
\begin{equation}
{{1}\over{2}}m\overline{V^2}(t)=KT(t),
\label{eq4}
\end{equation}
where $K$ is the Boltzmann constant and $\overline{V^2}(t)$ corresponds to the 
squared average velocity averaged over the ensemble of particles. The knowledge 
of $\overline{V^2}(t)$ directly gives the temperature $T(t)$.

We know from the thermodynamics \cite{blu} that an amount of heat 
transferred in a process depends on the temperature \footnote{This is true if 
no phase transitions are in course.} $dQ=c dT$, where $dQ$ is an infinitesimal 
amount of heat transferred at the price of an infinitesimal variation $dT$ in 
the temperature. The parameter $c$ corresponds to the heat capacity of the gas 
of particles. For an ideal gas $c=KN_p$ where $N_p$ is the total number of 
particles in the gas \cite{loskutov}. With these we have the left hand side of 
Eq. (\ref{eq3}) is written as ${{\partial Q}\over{\partial 
t}}={{cm}\over{2K}}{{\partial}\over{\partial t}}\overline{V^2}(t)$. The next 
step is to obtain the expression of the right side of Eq. (\ref{eq3}). Since the 
temperature gradient can only happen along the collision zone, we can consider 
an approximation that
\begin{equation}
{{\partial T}\over{\partial x}}\cong{{\Delta T}\over{\Delta 
x}}={{T-T_b}\over{\Delta x}},
\label{eq5}
\end{equation}
where $\Delta x$ is measured along the collision zone. To obtain $\Delta x$ we 
note that the radius of the boundary can assume two extrema: 
$R_{max}=1+\eta(1+\varepsilon)\cos(p\theta)$ and 
$R_{min}=1+\eta(1-\varepsilon)\cos(p\theta)$, where $R_{max}$ and $R_{min}$ 
correspond to the maximum and minimum values of the radius when time varies. 
The collision zone then is a region given by $\Delta 
R=R_{max}-R_{min}=2\eta\varepsilon\cos(p\theta)$. We see that $\Delta R$ is not 
constant being dependent directly on $\theta$ and has the property that 
$\overline{\Delta R}=0$. Therefore, an approximation for $\Delta x$ is obtained 
from $\Delta x=\sqrt{\overline{(\Delta R)^2}}$ where
\begin{equation}
\overline{(\Delta 
R)^2}={{1}\over{2\pi}}\int_0^{2\pi}4\eta^2\varepsilon^2\cos^2(p\theta)d\theta.
\label{average}
\end{equation}
A straightforward calculation gives $\Delta x=\sqrt{2}\eta\varepsilon$. Hence the 
expression of ${{\Delta T}\over{\Delta 
x}}={{T-T_b}\over{\sqrt{2}\eta\varepsilon}}$. Incorporating these approximations in 
the heat transfer equation we end up with
\begin{equation}
{{cm}\over{2K}}{{\partial}\over{\partial t}}{\overline{V^2}}=-{{\kappa 
\ell}\over{\sqrt{2}\eta\varepsilon}}\left[{{m}\over{2K}}{\overline{V^2}}
-T_b\right ] .
\label{eq6}
\end{equation}

Equation (\ref{eq6}) is a first order differential equation and that when 
solved properly leads to the following result
\begin{equation}
\overline{V^2}(t)={{2K}\over{m}}T_b+\left[V_0^2-{{2K}\over{m}}T_b\right]e^{-{{2 
\pi\kappa} \over{\sqrt{2}\eta\epsilon c}}t}.
\label{eq7}
\end{equation}
From the energy equipartition theorem, the temperature is written as
\begin{equation}
T(t)=T_b+\left[T_0-T_b\right]e^{-{{2\pi\kappa} \over{\sqrt{2}\eta\epsilon c}}t}.
\label{eq8}
\end{equation}

Let us now discuss some possibilities to study from experimental approach. 
Suppose a gas of particles is injected in the billiard with a low initial 
velocity such that  $T_0\gg T_b$. From Eq. (\ref{eq8}) and considering only the 
dominant term we have
\begin{equation}
T(t)\cong T_b+T_0e^{-{{2\pi\kappa} \over{\sqrt{2}\eta\epsilon c}}t},
\label{eq9}
\end{equation}
therefore, confirming an exponential decay for short $t$ and a convergence to 
the stationary state at $T(t)=T_b$ when $t\rightarrow \infty$. Other type of 
behavior is observed when an ensemble of particles is injected in the billiard 
with very low energy such that $T_0\ll T_b$. Expanding the exponential in 
Taylor series and keeping only the dominant term we end up with
\begin{equation}
T(t)=T_b{{{2\pi\kappa} \over{\sqrt{2}\eta\epsilon c}}t}.
\label{eq10}
\end{equation}
This result confirms the temperature grows at short time linearly in time hence 
leading the average velocity $\overline{V}(t)=\sqrt{\overline{V^2}}$ to grow 
with square root of time, hence
\begin{equation}
\overline{V}(t)=\sqrt{T_b{{{4\pi K\kappa} \over{\sqrt{2}m\eta\epsilon 
c}}}}\sqrt{t}.
\label{eq11}
\end{equation}

\section{Billiards approach}
\label{sec3}

\begin{figure}[htb]
\centerline{\includegraphics[width=1\linewidth]{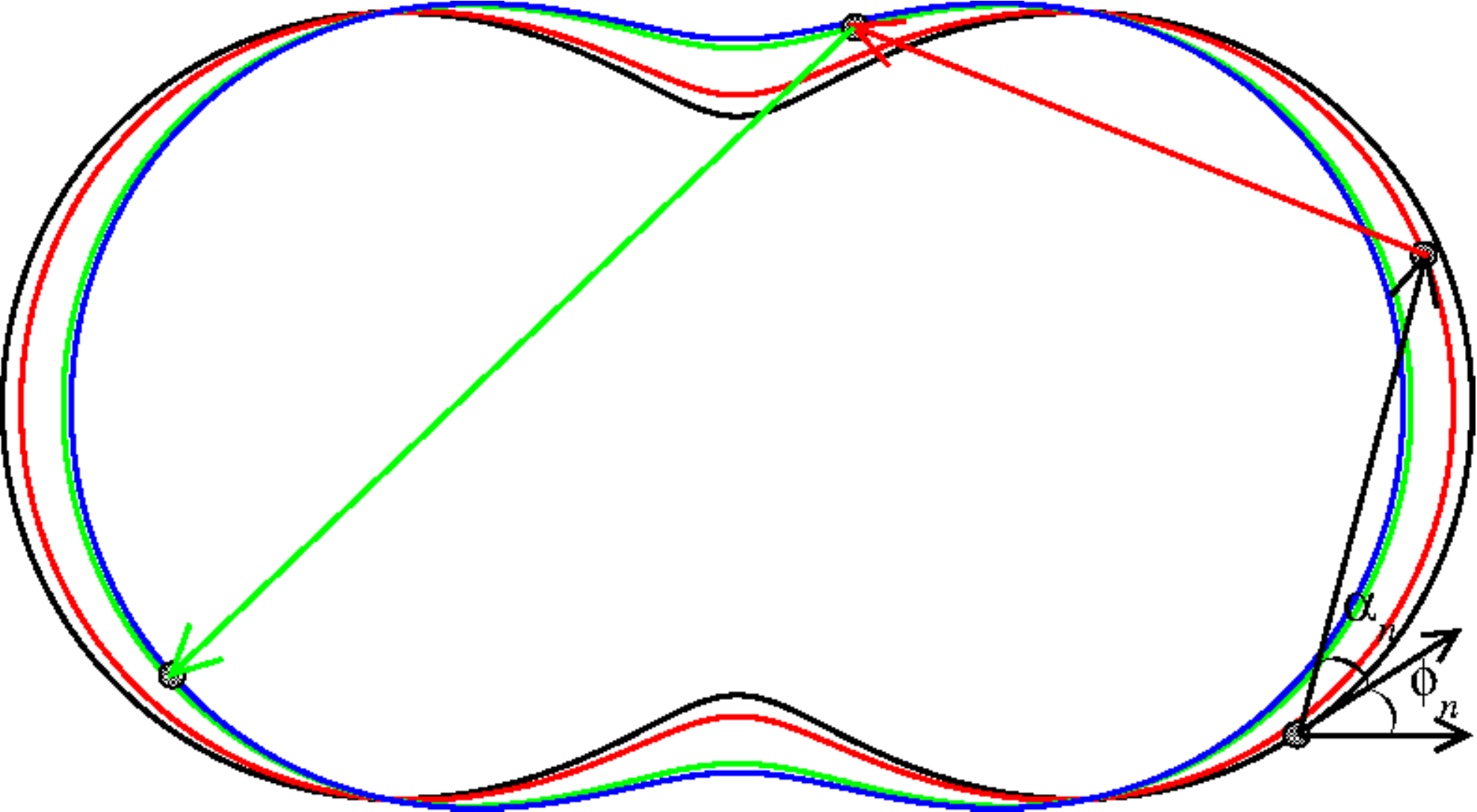}}
\caption{Illustration of four snapshots of the boundary at the four collisions.}
\label{Fig2}
\end{figure}

We now discuss how to construct the equations of the mapping that describe the 
dynamics of the particle inside of the billiard. The mapping gives the angular 
position of the particle $\theta$, the angle that the trajectory of the particle 
forms with a tangent line at the position of the collision $\alpha$, the 
absolute velocity of the particle $|\vec{V}|$ and finally the instant of the 
collision with the boundary $t$ at the impact $n^{th}$ with the further impact 
$(n+1)^{th}$. Figure \ref{Fig2} shows a typical illustration of a billiard and 
the angles used to describe the dynamics of the model.

The position of the particle at a given state 
$(\theta_n,\alpha_n,|\vec{V}_n|,t_n)$, written as a function of time is
\begin{eqnarray}
X(t)&=&X(\theta_n,t_n)+|\vec{V_n}|\cos(\alpha_n+\phi_n)(t-t_n),\\
Y(t)&=&Y(\theta_n,t_n)+|\vec{V_n}|\sin(\alpha_n+\phi_n)(t-t_n),
\end{eqnarray}
where the time $t\ge t_n$ with $X(\theta_n,t_n)=R(\theta_n,t_n)\cos(\theta_n)$ 
and $Y(\theta_n,t_n)=R(\theta_n,t_n)\sin(\theta_n)$. As soon as the $\theta$ is 
known, the angle $\phi$, which corresponds to the angle between the tangent 
line and the horizontal at $X(\theta),Y(\theta)$ is 
$\phi=\arctan[Y^{\prime}(\theta,t)/X^{\prime}(\theta,t)]$ where 
$Y^{\prime}(\theta,t)=dY/d\theta$ and $X^{\prime}(\theta,t)=dX/d\theta$.  

Considering the particle travels with a constant speed between collisions, 
the distance traveled by the particle measured with respect to the origin of 
the coordinate system is given by $R_p(t)=\sqrt{X^2(t)+Y^2(t)}$. The angular 
position $\theta_{n+1}$ is obtained by solving the equation 
$R_p(\theta_{n+1},t_{n+1})=R(\theta_{n+1},t_{n+1})$. The time at collision $n+1$ 
is given by
\begin{eqnarray}
t_{n+1}=t_n+{{\sqrt{\Delta X^2+\Delta Y^2}} \over \vert\vec{V_n}\vert},
\label{eqz}
\end{eqnarray}
where $\Delta X=X_p(\theta_{n+1},t_{n+1})-X(\theta_{n},t_{n})$ and $\Delta 
Y=Y_p(\theta_{n+1},t_{n+1})-Y(\theta_{n},t_{n})$. 

We notice that the referential frame of the boundary is non inertial. We 
assume also the collisions of the particle with the boundary are inelastic, 
hence there is a fractional loss of energy upon collision, which we consider  
only with respect to the normal component of the velocity. Then at the instant 
of collision the reflection laws are
\begin{eqnarray}
\vec{V^{\prime}}_{n+1}\cdot\vec{T}_{n+1}&=&\vec{V^{\prime}}_n\cdot 
\vec{T}_{n+1},\label{law1}\\
\vec{V^{\prime}}_{n+1}\cdot\vec{N}_{n+1}&=&-\gamma\vec{V^{\prime}}_n\cdot 
\vec{N}_{n+1},\label{law2}
\end{eqnarray}
where the unit tangent and normal vectors are
\begin{eqnarray}
\vec{T}_{n+1}&=&\cos(\phi_{n+1})\hat{i}+\sin(\phi_{n+1})\hat{j},\\
\vec{N}_{n+1}&=&-\sin(\phi_{n+1})\hat{i}+\cos(\phi_{n+1})\hat{j},
\end{eqnarray}
Here $\gamma\in[0,1]$ is the restitution coefficient. If $\gamma=1$ we have 
completely elastic collisions while $\gamma<1$ leads the particle to experience 
a partial loss of velocity upon collisions. The term $\vec{V^{\prime}}$ 
corresponds the velocity of the particle measured in the non-inertial reference 
frame. We can then obtain the tangential and normal components of the velocity 
after collision $n+1$ as
\begin{eqnarray}
\vec{V}_{n+1}\cdot\vec{T}_{n+1}&=&\vec{V}_{n}\cdot\vec{T}_{n+1},\\
\vec{V}_{n+1}\cdot\vec{N}_{n+1}
&=&-\gamma\vec{V}_{n}\cdot\vec{N}_{n+1}+\nonumber\\
&+&(1+\gamma)\vec{V}_{b}(t_{n+1}+Z(n)
)\cdot\vec{N}_{n+1},
\end{eqnarray}
where $\vec{V}_{b}(t_{n+1}+Z(n))$ denotes the velocity of the boundary that is 
given by
\begin{eqnarray}
\vec{V}_{b}(t_{n+1})={dR(t)\over {dt}}\Big\vert_{t_{n+1}}
[\cos(\theta_{n+1})\widehat{i}+\sin(\theta_{n+1})\widehat{j}],
\end{eqnarray}
and $Z(n)\in[0,2\pi]$ is a random number introduced in the argument of the 
velocity of the moving wall to simulate stochasticity into the model.

Finally, the  velocity of the particle after the collision $(n+1)$ is given by
\begin{eqnarray}
|\vec{V}_{n+1}|=\sqrt{(\vec{V}_{n+1}\cdot\vec{T}_{n+1}
)^2+(\vec{V}_{n+1}\cdot\vec{N}_{n+1})^2},
\label{eq012}
\end{eqnarray}
when the angle $\alpha_{n+1}$ is written as
\begin{eqnarray}
\alpha_{n+1}=\arctan
\left[{\vec{V}_{n+1}\cdot\vec{N}_{n+1} \over
\vec{V}_{n+1}\cdot\vec{T}_{n+1}} \right]~.
\end{eqnarray}

With the equations above we can now discuss some of the statistical properties 
for the average velocity of the particle.

\subsection{Stationary state}

To investigate the average velocity of an ensemble of particles we make the 
following assumption. We consider the probability distribution for the velocity 
in the two-dimensional phase space $\alpha~vs.~\theta$ is uniform. In the 
stochastic model, the one which gives random numbers $Z$ in the argument of 
the velocity of the moving wall at each collision, this is observed. If we take 
the expression of $|\vec{V}_{n+1}|$ and average the squared velocity for the 
ranges $\theta\in[0,2\pi]$, $\alpha\in[0,\pi]$ and 
$t\in[0,2\pi]$ we end up with
\begin{equation}
\overline{V^{2}}_{n+1}={{\overline{V^2}_n}\over{2}}+
{{\gamma^2\overline{V^2}_n}\over{{2}}}+{{(1+\gamma)^{2}\eta^{2}\varepsilon^{2}}
\over{8}}. 
\label{eq_v2}
\end{equation} 

In the steady state regime the mean-squared velocity is obtained considering
$\overline{V^{2}_{n+1}}=\overline{V^{2}_{n}}=\overline{V^{2}}$, and after 
isolating $\overline{V^{2}}$ we obtain
\begin{eqnarray}
\overline{V^{2}}={{(1+\gamma)\eta^{2}\varepsilon^{2}}\over{4(1-\gamma)}}.
\end{eqnarray}
If we define the root mean square velocity as 
$\overline{V}=\sqrt{\overline{V^{2}}}$, we have
\begin{equation}
\overline{V}={{\eta\varepsilon}\over{2}}\sqrt{(1+\gamma)}(1-\gamma)^{-1/2}.
\label{eq_sat}
\end{equation}

We notice from Eq. (\ref{eq_sat}) that the exponent heading the term 
$(1-\gamma)$ is $-1/2$ while the exponent heading the parameters 
$(\eta\varepsilon)$ is $1$. We discuss these exponents latter on.

\subsection{Dynamical regime}

An easy way to study the dynamical regime is transform the difference equation 
given in Eq. (\ref{eq_v2}) into a differential equation where the 
solutions can be easier to track. We assume for a large ensemble that
\begin{equation}
\overline{V^{2}}_{n+1}-\overline{V^{2}_{n}}={{\overline{V^{2}}_{n+1}-\overline{
V^{2}_{n}}}\over{(n+1)-n}}\cong {d{\overline{V^{2}}}\over{dn}},
\end{equation}
which leads to
\begin{equation}
{d{\overline{V^{2}}}\over{dn}}={{\overline{V^{2}}}\over{2}}(\gamma^{2}-1)+{
{(1+\gamma)^{2}\eta^{2}\varepsilon^{2}}\over{8}}.
\end{equation}
A straightforward integration considering the initial condition $V_0$ at $n=0$ 
gives
\begin{equation}
\overline{V^{2}}(n)=\overline{V^{2}_{0}}e^{{{(\gamma^2-1)}\over{2}}n}+ 
{{(1+\gamma)}\over{4(1-\gamma)}}\eta^{2}\varepsilon^{2}\bigg[1-e^{{{(\gamma^{2}
-1)}\over{2}}n}\bigg].
\label{v_din}
\end{equation}

The dynamics of $\overline{V}(n)=\sqrt{\overline{V^2}(n)}$ is described by
\begin{equation}
\overline{V}(n)=\sqrt{\overline{V^{2}_{0}}e^{{{(\gamma^2-1)}\over{2}}n}
+{{(1+\gamma)}\over{4(1-\gamma)}}\eta^{2}\varepsilon^{2}\bigg[1-e^{{{(\gamma^{2}
-1)}\over{2}}n}\bigg]}.
\label{v_full}
\end{equation}

Two important limits are obvious from Eq. (\ref{v_full}). The first one 
considered is when $V_0\gg {{(1+\gamma)^{1/2}}\over{2}}(1-\gamma)^{-1/2}
\eta\varepsilon$ hence leading to an exponential decay of the velocity
\begin{equation}
\overline{V}(n)\cong V_0e^{{{(\gamma^2-1)}\over{4}}n}\cong 
V_0e^{{{(\gamma-1)}\over{2}}n}.
\label{v_decay}
\end{equation}
The second one is observed when the initial velocity is sufficiently small, say 
$V_0\cong 0$, the dominant expression for $\overline{V}(n)$ is
\begin{equation}
\overline{V}(n)={{(1+\gamma)^{1/2}}\over{2}}(1-\gamma)^{-1/2}
\eta\varepsilon 
\bigg[1-e^{{{(\gamma^{2}-1)}\over{2}}n}\bigg]^{1/2}.
\label{inicial}
\end{equation} 

A Taylor expansion in Eq. (\ref{inicial}) gives that
\begin{equation}
\overline{V}(n)\sim \eta\varepsilon \sqrt{n}.
\label{v_ini}
\end{equation}

\subsection{Numerical simulations}

\begin{figure}[t]
\centerline{\includegraphics[width=1\linewidth]{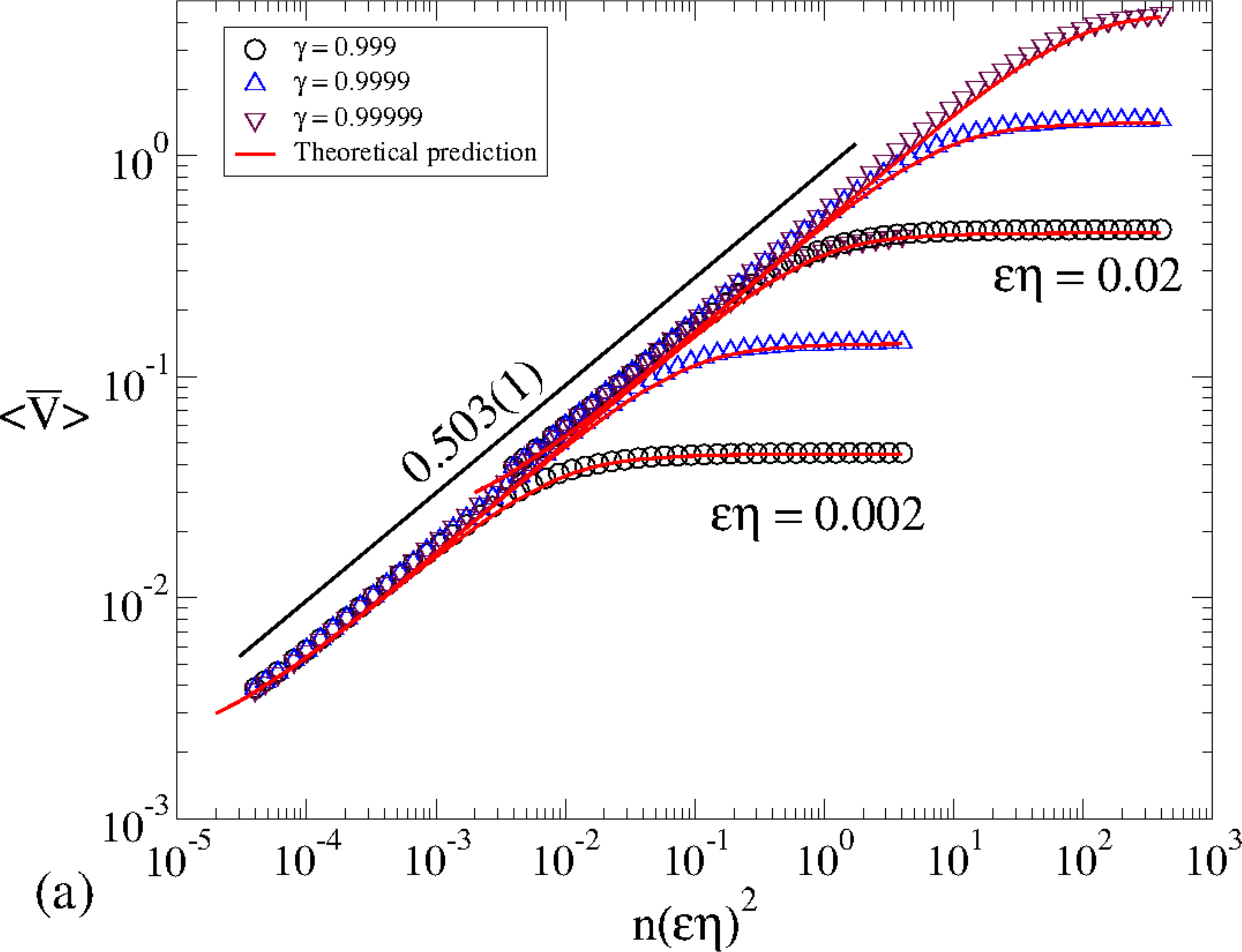}}
\centerline{\includegraphics[width=1\linewidth]{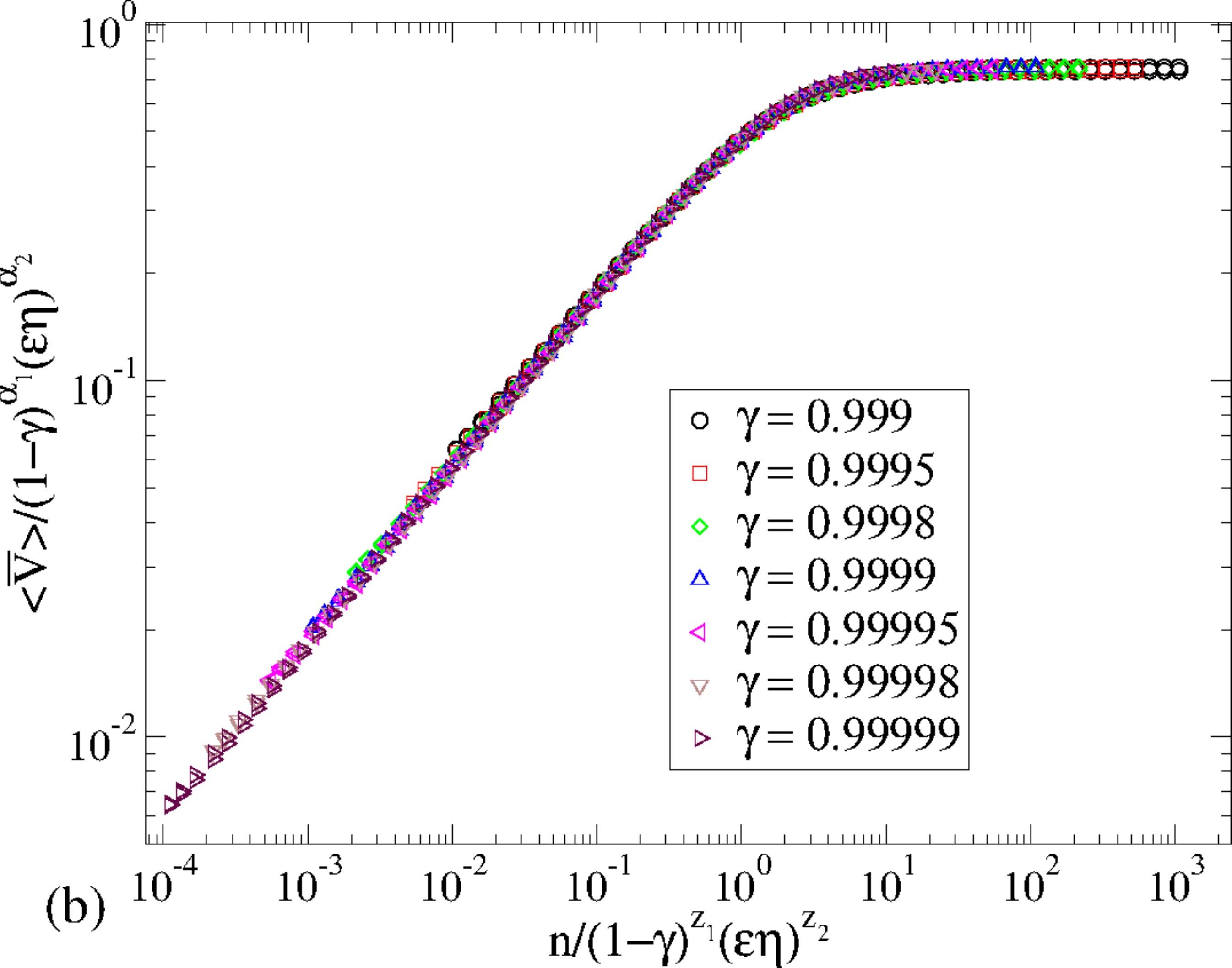}}
\caption{(a) Plot of $<\overline{V}>~vs.~n$ for different values of 
$\gamma$ and two combinations of $\eta\varepsilon$. (b) Overlap of the curves 
shown in (a) onto a single and universal plot after the following scaling 
transformations: $n\rightarrow n/[(1-\gamma)^{z_1}(\eta\varepsilon)^{z_2}]$ and 
$<\overline{V}>\rightarrow<\overline{V}>/[(1-\gamma)^{\alpha_1}(\eta\varepsilon)^{
\alpha_2}]$. The straight line gives the theoretical prediction [Eq. 
(\ref{vrms})].}
\label{Fig3}
\end{figure}

Let us now discuss the behavior of the squared average velocity via numerical 
simulations. The range of $\gamma$ we are interested in is $\gamma\rightarrow 
1$ therefore close to the transition from conservative to dissipative dynamics. 
According to the LRA conjecture, if $\gamma=1$ (conservative case) the average 
velocity must grow unbounded. However for $0<\gamma<1$ there must exist a limit 
for the growth, as foreseen for the two previous sections. The transition is 
better characterized for $(1-\gamma)$. The simulations were made in such a 
way that each initial condition has a fixed initial velocity, $V_0=10^{-3}$, 
$\eta\varepsilon \in [0.002,0.02]$ and randomly chosen $\alpha_0 \in [0,\pi]$, 
$\theta_0 \in [0,2\pi]$, $t_0 \in [0,2\pi]$. Moreover, after each time step, a 
random number $[Z(n)]$ is drawn in the equation of the velocity of the moving 
wall introducing stochasticity 
into the model. For computing the average velocity numerically, two different 
procedures were applied: (i) we evaluate the average velocity over the 
orbit for a single initial condition and; (ii) average the velocity over an 
ensemble of initial conditions. Hence, the average velocity is written as
\begin{eqnarray}
<\overline{V}>(n)={{1}\over{M}}\sum_{i=1}^M{{1}\over{n+1}}\sum_{j=0}^nV_{i,j}~,
\label{eq16}
\end{eqnarray}
where the index $i$ corresponds to a sample of an ensemble of initial 
conditions and $M=2000$ denotes the number of different initial conditions. A 
plot of $<\overline{V}>~vs.~n$ for different values of $\gamma$ is shown in Fig. 
\ref{Fig3} (a). 

From Fig. \ref{Fig3}(a) we see two different kinds of behaviors. At small $n$, 
the average velocity grows to start with to a power law and eventually it bends 
towards a regime of saturation for large $n$. The change from growth to the 
saturation is given by a characteristic crossover $n_x$. We notice 
that a transformation $n \rightarrow n(\eta\varepsilon)^2$ coalesces all curves 
to grow together before moving to the saturation. The behavior shown in Fig. 
\ref{Fig3}(a) allows us to propose that: (i) For short $n$, say $n\ll n_{x}$, 
the growth regime is described by 
$<\overline{V}>\propto[(\eta\varepsilon)^{2}n]^{\beta}$ where $\beta$ is the 
acceleration exponent; (ii) For large enough $n$, typically $n \gg n_{x}$ we 
have 
$<\overline{V}_{sat}>\propto(1-\gamma)^{\alpha_{1}}(\eta\varepsilon)^{\alpha_{2 
}}$ where $\alpha_{1}$ and $\alpha_{2}$ are the saturation exponents; (iii) The 
crossover $n_x$ that marks the changeover from growth to the saturation is given 
by $n_{x}\propto(1-\gamma)^{z_{1}}(\eta\varepsilon)^{z_{2}}$ where $z_{1}$ and 
$z_{2}$ are crossover exponents.

The three previous hypotheses allow us to describe the behavior of 
$<\overline{V}>$ by a homogeneous function of the type 
\begin{eqnarray}
<\overline{V}>&[&(\eta\varepsilon)^{2}n,\eta\varepsilon,(1-\gamma)~]
=\nonumber\\
&&l<\overline { V }
>[l^{a}(\eta\varepsilon)^{2}n,l^{b}\eta\varepsilon,l^{d}(1-\gamma)],
\end{eqnarray}
where $l$ is a scale factor, $a$, $b$ and $d$ are characteristic exponents that 
in principle must be related to the scaling exponents. A straightforward 
calculation gives the two scaling laws
\begin{eqnarray}
z_{1}={{\alpha_{1}}\over{\beta}},~~~~~~~~ 
z_{2}={{\alpha_{2}}\over{\beta}}-2.\label{sl_2}
\end{eqnarray}

\begin{figure}[t]
\centerline{\includegraphics[width=1\linewidth]{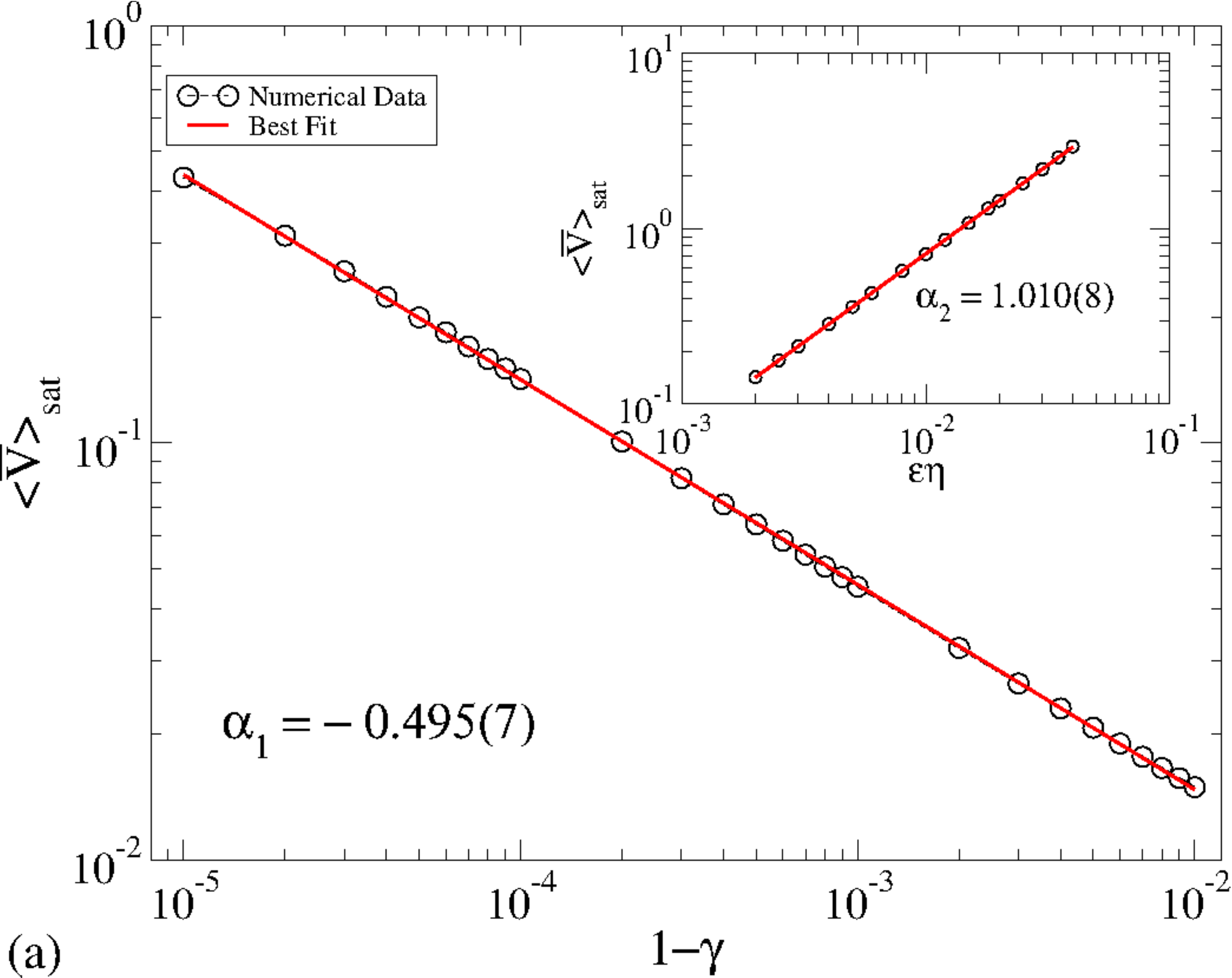}}
\centerline{\includegraphics[width=1\linewidth]{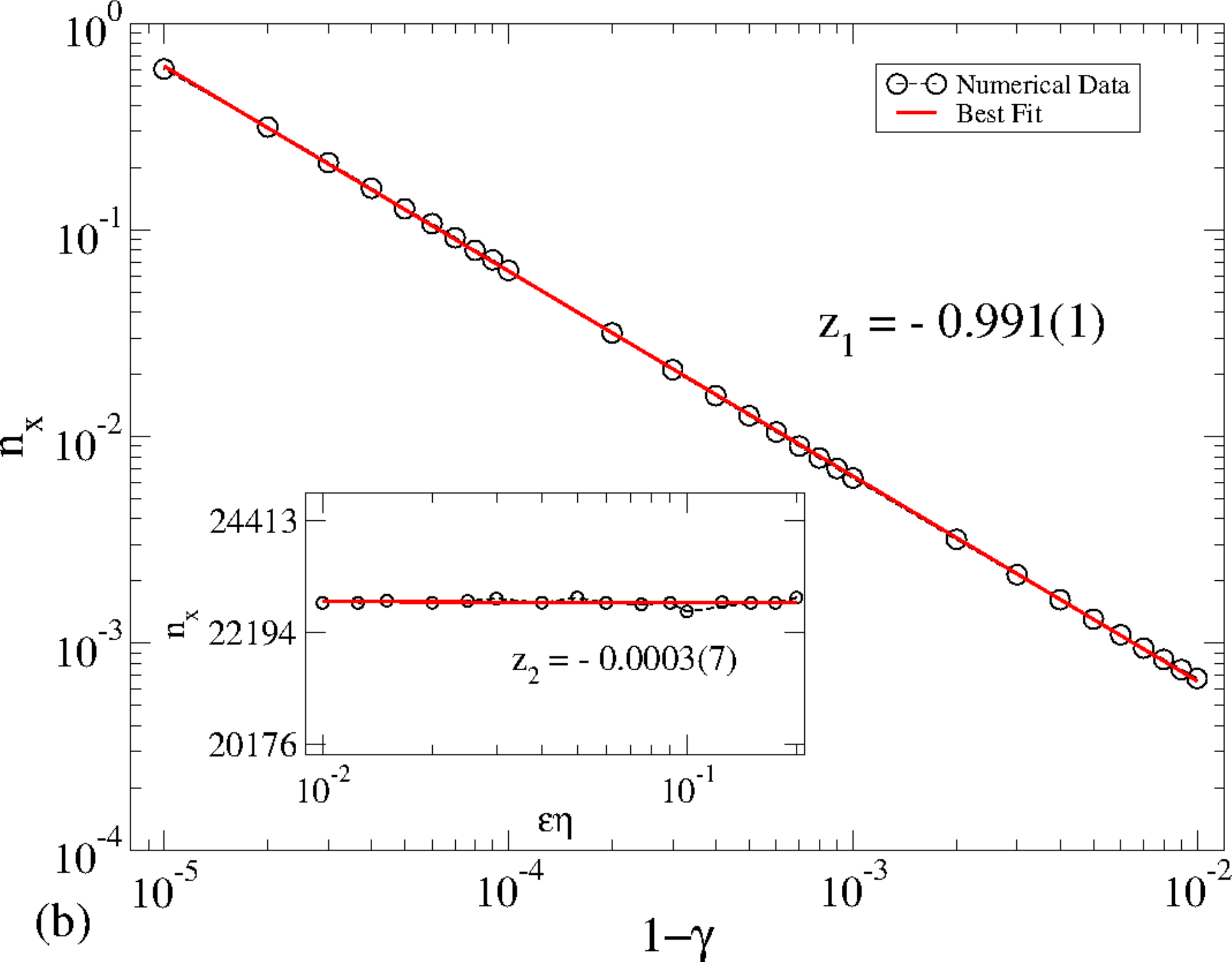}}
\caption{Behavior of: (a) $<\overline{V}_{sat}>$ and (b) $n_x$ as a function of 
$(1-\gamma$). The inset shown the behavior of $<\overline{V}_{sat}>$ and $n_x$ 
for different values of $\varepsilon\eta$.}
\label{Fig4}
\end{figure}

All the five exponents can be obtained numerically. A power law fitting in the 
regime of growth for $<\overline{V}>$ gives $\beta=0.503(1)\simeq 1/2$. 
Similar values were obtained for all curves we simulated for the range of 
$\gamma\in[0.999,0.99999]$. If we keep fixed $\eta\varepsilon$ and vary 
$\gamma$, a power law fitting for $<\overline{V}_{sat}>~vs.~ (1-\gamma)$ 
furnishes $\alpha_1=-0.495(7)\cong-1/2$, as shown in Fig. \ref{Fig4} (a). A 
fitting to the plot of ${n}_{x}~vs.~(1-\gamma)$ gives $z_1=-0.991(1)\cong-1$. 
Finally if we keep constant $(1-\gamma)$ a fitting to the behavior of 
$<\overline{V}_{sat}>~vs.~\eta\varepsilon$ gives $\alpha_1=1.010(8)\cong 1$ 
while a plot of $n_x~vs.~\eta\varepsilon$ yields $z_2=-0.0003(7)\cong0$. When 
the two scaling laws (\ref{sl_2}) are used to check the exponents, the results 
obtained are remarkably in well agreement with the simulations.

\subsection{Averaging the velocity along $n$}

As given by Eq. (\ref{v_din}), the squared velocity was obtained considering an 
average over an ensemble of particles. However, the simulations were made using 
either ensemble average as well as average on time. Therefore, we have to find a 
corresponding expression of the squared velocity when it is also averaged over 
the number of collisions $n$. The average squared velocity is written as
\begin{equation}
<\overline{V^2}(n)>={{1}\over{n+1}}\sum_{i=0}^n\overline{V^2}(i).
\end{equation}
The summation over the exponential terms converges since their arguments are 
negative. The convergence of the exponential terms is
\begin{equation}
\sum_{i=0}^ne^{({{\gamma^2-1}\over{2}})i}=\left[{{1-e^{({{\gamma^2-1}\over{2}})
(n+1)}}\over{1-e^{{\gamma^2-1}\over{2}}}}\right],
\end{equation}
hence the root mean squared velocity is written as 
$V_{rms}(n)=\sqrt{<\overline{V^2}(n)>}$, therefore
\begin{widetext}
\begin{equation}
V_{rms}(n)=\sqrt{{{(1+\gamma)\eta^2\varepsilon^2}\over{4(1-\gamma)}}+{{1}\over{
(n+1) } } 
\left[V_0^2-{{(1+\gamma)\eta^2\varepsilon^2}\over{4(1-\gamma)}}\right]\left[{{1-e^{ 
{(n+1){{(\gamma^2-1)}\over{2}}}}}\over{1-e^{{{(\gamma^2-1)}\over{2}}}}}\right]}.
\label{vrms}
\end{equation}
\end{widetext}
A plot of Eq. (\ref{vrms}) is represented as a continuous line in Fig. 
\ref{Fig3}(a).

Two important limits for Eq. (\ref{vrms}) are:
\begin{enumerate}
\item{$n=0$, that leads to $V_{rms}(0)=V_0$;
}
\item{Considering the limit of $n\rightarrow \infty$, we have
\begin{equation}
V_{rms}(n\rightarrow\infty)=\sqrt{{{(1+\gamma)\eta^2\varepsilon^2}\over{
4(1-\gamma)}}}.
\label{v_sat}
\end{equation}
}
\end{enumerate}

With Eq. (\ref{vrms}) we can discuss the behavior of $V_{rms}$ for short $n$. 
In the limit of $\gamma\approx 1$ we can Taylor expand the two exponentials of 
Eq. (\ref{vrms}). Because of term $(n+1)$ in the denominator of Eq. 
(\ref{vrms}), the expansion of the exponential of the numerator must go until 
second order while the denominator can go only to the first. Grouping the terms 
properly we obtain the expression of $V_{rms}(n)$ when $V_0\cong0$ as
\begin{equation}
V_{rms}(n)\cong{{(1+\gamma)\eta\varepsilon}\over{4}}\sqrt{(n+1)}.
\label{vgrowth}
\end{equation}
When $n\gg 1$ such that $\sqrt{(n+1)}\cong \sqrt{n}$ then we have 
$V_{rms}(n)\cong{{(1+\gamma)\eta\varepsilon}\over{4}}\sqrt{n}$.

\subsection{Critical exponents}

The five relevant critical exponents that describe the scaling properties of 
the average velocity curves are $\beta$, $\alpha_i$ and $z_i$ with $i=1,2$. 
The exponents $\alpha_1$ and $\alpha_2$ are obtained for the regime of 
$n\rightarrow\infty$. From Eq. (\ref{v_sat}) we obtain $\alpha_1=-1/2$ and 
$\alpha_2=1$. The exponent $\beta$ comes from Eq. (\ref{vgrowth}). When $n\gg 
1$ we have $\beta=1/2$. Finally, the crossover iteration number $n_x$ can be 
estimated when Eq. (\ref{vgrowth}) intersects  Eq. (\ref{v_sat}). A 
straightforward calculation gives
\begin{equation}
n_{x}={{4}\over{(1+\gamma)}}(1-\gamma)^{-1}.
\end{equation}
Then we conclude that $z_1=-1$ and $z_2=0$.

\subsection{Velocity distribution}

\begin{figure}[htb]
\centerline{\includegraphics[width=1\linewidth]{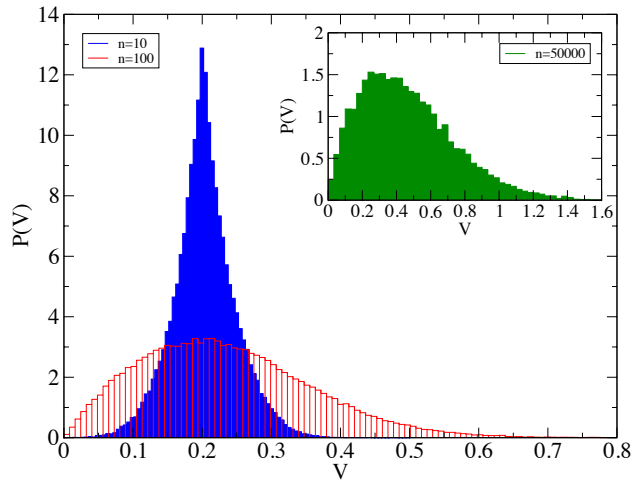}}
\caption{Plot of the normalized probability distribution for the velocity for 
an ensemble of $10^5$ particles in the dissipative and stochastic oval 
billiard. Blue (dark gray) was obtained after $10$ collisions while red bars
(light gray) was obtained after $100$ collisions. The in-box was obtained after 
$50,000$ collisions. The initial velocity used was $V_0=0.2$ and control 
parameters $\epsilon\eta=0.02$ and $\gamma=0.999$ for $p=2$.}
\label{distribution}
\end{figure}

Let us discuss here how is the shape of the velocity distribution for the 
dynamics in the dissipative case. It is important to notice that the lowest 
velocity for a moving particle is limited to the lowest velocity of the moving 
boundary, hence $V_l=-\eta\epsilon$. Upward velocities are unbounded but 
unlimited energy growth is not observed due to the dissipation. The lower limit 
for the velocity plays a major rule on the distribution of the velocity and to 
illustrate this we discuss the following case. Suppose an ensemble of initial 
conditions with different angular variables, $\alpha$, $\theta$, but with the 
same initial velocity is given. The initial velocity is chosen in such a 
way that it is located in a region above of lower velocity limit and, at the 
same time, below than the saturation. The dynamics evolves as follows. For 
short number of collision with the boundary, part of the ensemble of particles 
raises the velocity while the other part reduces velocity. This distribution is 
Gaussian, as shown in Fig. \ref{distribution} in blue (dark gray) color for and 
initial velocity of $V_0=0.2$ and distribution collected after $10$ collisions 
with the boundary.
The parameters used were $\epsilon\eta=0.02$, $\gamma=0.999$ and $p=2$, 
although other values would lead to similar results. Moreover, a total of 
$2.5\times10^6$ different initial conditions were considered in the ensemble. As soon as 
the dynamics evolves, the Gaussian distribution flats itself in both sides 
until the left hand side curves touches the lower limit of the velocity. See the 
red (light gray) bars obtained for the distribution after $100$ collisions with 
the boundary. At this point, the distribution experiences a break of symmetry 
and hence if the initial velocity of the distribution is lower that the 
saturation, the average velocity starts to grow until approaches the saturation. 
From this point of symmetry break and beyond this point, the distribution is not Gaussian 
anymore and it has similar shape as shown in the in-box of Fig. \ref{distribution}. 
Such distribution was obtained after $50,000$ collisions of the ensemble of 
particles with the boundary. Although the distribution is not Gaussian anymore 
due to the break of symmetry at $V=V_l$, the distribution has clearly a peak and 
decays monotonically for large enough values of velocity warranting convergence 
for the average velocity as well other momenta of the distribution. It is worth to mention that such a break in the symmetry in the probability distribution was previously observed for a one dimensional Fermi Ulam model \cite{physicaa}. There, the authors show that the velocity/energy distribution can be described perfectly by a folded normal distribution.

\section{Connection between the two approaches}
\label{sec4}

The results discussed in Section \ref{sec2} involving the heat transfer equation 
were obtained as a function of the time $t$ while in Section \ref{sec3} the 
results were discussed using the number of collisions $n$. It is important to 
mention that the time $t$ and the number of collisions $n$ are variables not 
trivially connected with each other. It happens because a particle moving with 
high speed can experience many more collisions with the boundary when compared with 
a particle with low energy at the same interval of time. In this section we 
discuss a way to make a connection of the two variables therefore linking the 
results discussed in Secs. \ref{sec2} and \ref{sec3}.

Given the particle travels with constant velocity between collisions, the 
length of time between two collisions is $\Delta t=d/|\vec{V}|$ where $d$ is 
the distance traveled by the particle and $|\vec{V}|$ is its absolute velocity. 
Therefore, the total time spent at $n$ collisions is written as
\begin{equation}
\tau=\sum_{i=0}^n{{d_i}\over{|\vec{V}_i|}}.
\label{tau}
\end{equation}
The summation in Eq. (\ref{tau}) seems to be not easy to be made. As an attempt 
to have an explicit expression involving the relevant parameters of the system 
considered we will do the summation in two stages evaluating then the numerator 
separately then the denominator.

From the numerator we can estimate the mean free path, which we represent as
\begin{equation}
\overline{d}={{1}\over{(n+1)}}\sum_{i=0}^nd_i,
\end{equation}
where $d_i$ is defined as the distance from two collisions as
\begin{equation}
d_i=\sqrt{[x(\theta_{i+1})-x(\theta_i)]^2+[y(\theta_{i+1})-y(\theta_i)]^2},
\end{equation}
where $x(\theta)=R(\theta)\cos(\theta)$ and $y(\theta)=R(\theta)\sin(\theta)$. 
The dynamics of each particle is chaotic, therefore when we do an average over 
$\theta\in[0,2\pi]$ we obtain
\begin{equation}
\overline{d}=\sqrt{2+\eta^2\left[1+{{\varepsilon^2}\over{2}}\right]}.
\label{dmedio}
\end{equation}

The second part we have to consider is $\sum_{i=0}^n{{1}\over{V_i}}$. To do 
that we consider the variation of the velocity from the collision $i$ to $(i+1)$ 
is small so that the summation can be approximated by
\begin{equation}
\sum_{i=0}^n{{1}\over{V_i}}\cong\int_{0}^n{{1}\over{V(n^{\prime})}}
dn^{\prime}.
\end{equation}
The expression for $\tau$ obtained for the explicit form of $\overline{V}$ as 
shown in Eq. (\ref{v_full}) is not an easy equation to deal with and the result 
is reported in the Appendix 2 for the interested reader. Instead of dealing 
with the whole equation we consider an easier approach. As discussed in Ref. 
\cite{diego}, the behavior of the average squared velocity, when settled in 
scaling variables can be described by a function of the type
\begin{equation}
f(x)=\left[{{x}\over{1+x}}\right]^{\beta},
\end{equation}
where $\beta$ is the accelerating exponent. In our case $\beta=1/2$. Therefore, 
the scaled variables considered are: 
$f\rightarrow{{\overline{V}\sqrt{(1-\gamma)}}\over{\eta\varepsilon}}$ and 
$x\rightarrow n(1-\gamma)$. Incorporating these two equations as the behavior 
of $\overline{V}(n)$ we end up with the following equation to be solved
\begin{equation}
\tau={{\overline{d}\sqrt{(1-\gamma)}}\over{\eta\varepsilon}}\int{{dn}\over{\sqrt{{
n(1-\gamma)}\over{1+n(1-\gamma)} } } },
\end{equation}
After doing the integration (see Appendix 3 for the result of the integral) 
and 
keeping only the leading term in $n$ we obtain
\begin{equation}
\tau\cong{{\overline{d}\sqrt{(1-\gamma)}}\over{\eta\varepsilon}}n.
\end{equation}

\section{Conclusions}
\label{sec5}
We have studied some dynamical and statistical properties of gas of non 
interacting particles in a time dependent and dissipative oval billiard. We have 
investigated the behavior of the average velocity of the particles as a 
function of time and the number of collisions with the moving boundary by using 
two different approaches, namely, involving  (i)  heat transfer and (ii) 
billiards. We have obtained an empirical expression for the average squared 
velocity by using the equilibrium condition at the steady state regime. Such an 
expression allowed us to make a connection with the thermodynamic, more 
precisely by using the  Fourier law for heat transfer. The resulting 
equation have   shown that the temperature of the gas reaches the thermal 
equilibrium for sufficiently long time. Our results also have demonstrated that  
the average squared velocity grows as a power law and after a crossover it tends 
to a constant plateau. Furthermore, the stronger is the dissipation the faster 
is the transition from growth to saturation. Finally, by using an empirical 
function to describe the behavior of the average squared velocity, we have 
shown that time and number of collisions are linearly correlated.

\section*{ACKNOWLEDGMENTS}
EDL thanks FAPESP (2012/23688-5) and CNPq (303707/2015-1) for financial support. 
MVCG acknowledges CNPq for financial support. DFMO thanks to James S. McDonnell 
Foundation.

\section*{Appendix 1}

This appendix is devoted to a short discussion on the heat flow equation 
\cite{blu}. The heat can indeed quantify an amount of energy which is 
transferred due to a temperature gradient. The amount of heat flowing along the 
temperature gradient depends on the thermal conductivity $\kappa$. The heat 
flows from a region of high to low temperature, therefore this flow is contrary 
to the temperature gradient. In a generic 3-D system, the heat flux vector 
$\vec{J}$ is written as $\vec{J}=-\kappa A \vec{\nabla} T$, where $A$ 
corresponds to a section of area perpendicular to where the flow of heat is 
flowing while $\vec{\nabla} T$ gives the gradient of temperature. The signal 
$(-)$ is introduced to represent a flow contrary to the temperature gradient, 
i.e. from higher to lower temperature. The vector $\vec{J}$ indeed represents a 
certain amount of energy which is flowing through an area $A$ at a given 
interval of time due to a gradient of temperature.

In the system we are considering in this paper, the flow of heat is not 
crossing a perpendicular area, but rather it crosses the border of the 
billiard. Hence, in the case 2-D as discussed, the heat transfer equation is 
written as 
\begin{equation}
J={{\partial Q}\over{\partial t}}=-\kappa \ell {{\partial T}\over{\partial 
x}},
\end{equation}
where $J$ represents the amount of heat which is transferred around the 
border $\ell$ of the billiard at a given instant of time due to a temperature 
gradient represented as ${{\partial T}\over{\partial x}}$. In our case, the 
thermal conductivity coefficient $\kappa$ denotes the constant of 
proportionality between the amount of energy flowing in the border of the 
billiard $\ell$ per unit of time due to a temperature gradient.

\section*{Appendix 2}

When we consider $V(n)$ as given by Eq. (\ref{v_full}) to obtain the 
expression for $\tau$ the direct integral is
\begin{equation}
\tau=\overline{d}\int_{0}^n{{dn^{\prime}}\over{\sqrt{\overline{V^{2}_{0}}e^{{{
(\gamma^2-1)}\over{2}}n^{\prime}}+{{(1+\gamma)}\over{4(1-\gamma)}}\eta^{2}
\varepsilon^{2}\bigg[1-e^{{{(\gamma^{2}-1)}\over{2}}n^{\prime}}\bigg]}}}.
\end{equation}

A straight integration yields
\begin{widetext}
\begin{equation}
\tau={{8\sqrt{2+\eta^2\left(1+{{\varepsilon^2}\over{2}}\right)}}\over{
\eta\varepsilon\sqrt { (1+\gamma) } (1-\gamma^2)}}\times\left[{\rm 
arctanh}\left({{\sqrt{{{(1+\gamma)\eta^2\varepsilon^2}\over{4(1-\gamma)}}
+\left(V_0^2-{{(1+\gamma)\eta^2\varepsilon^2}\over{4(1-\gamma)}}\right)e^{{{
(\gamma^2-1)}\over{2}}n}}}\over{{{\eta\varepsilon}\over{2}}\sqrt{{(1+\gamma)}\over{
(1-\gamma)}}}}\right)-{\rm arctanh}\left({{V_0}\over{{{\eta\varepsilon}\over{2}}
\sqrt{{{(1+\gamma)}\over{(1-\gamma)}}}}}\right)\right].\nonumber
\end{equation}
\end{widetext}
With some algebra one can isolate $n$ as a function of $\tau$ from the equation above.

\section*{Appendix 3}

The solution of the integral
\begin{equation}
\tau={{\overline{d}\sqrt{(1-\gamma)}}\over{\eta\varepsilon}}\int{{dn}\over{\sqrt{{
n(1-\gamma)}\over{1+n(1-\gamma)} } } },
\end{equation}
is given by
\begin{widetext}
\begin{equation}
\tau={{\overline{d}\sqrt{(1-\gamma)}}\over{\eta\varepsilon}}\left[
{{1}\over{2}}{{
2\sqrt{(n^2-n^2\gamma+n)}\sqrt{(1-\gamma)}
}\over{\sqrt{(1-\gamma)}\sqrt{-n(-1-n+n\gamma)}\sqrt{{{-n(1-\gamma)}\over{ 
-1-n+n\gamma } } } } }n\right]+
{{\overline{d}\sqrt{(1-\gamma)}}\over{\eta\varepsilon}}\left[{{1}\over{2}}{{
\ln\left[-{{1}\over{2}}\left({{-1-2n+2n\gamma-2\sqrt{(n^2-n^2\gamma+n)} 
\sqrt{1-\gamma}}\over { \sqrt { (1-\gamma) } } }\right)
\right]}\over{\sqrt{(1-\gamma)}\sqrt{-n(-1-n+n\gamma)}\sqrt{{{-n(1-\gamma)}\over
{-1-n+n\gamma}}}}}n\right].
\end{equation}
\end{widetext}

After grouping the terms and considering only the leading term we have
\begin{equation}
\tau={{\overline{d}\sqrt{(1-\gamma)}}\over{\eta\varepsilon}}\left[n\sqrt{1+{{1}
\over{n(1-\gamma)}} }~\right ].
\end{equation}
Expanding the square root and keeping only the first order we have
\begin{equation}
\tau={{\overline{d}\sqrt{(1-\gamma)}}\over{\eta\varepsilon}}\left[n+{{1}\over{
2( 1-\gamma)} } \right ],
\end{equation}
therefore
\begin{equation}
\tau\cong{{\overline{d}\sqrt{(1-\gamma)}}\over{\eta\varepsilon}}n.
\end{equation}



\end{document}